# Steganography- A Game of Hide and Seek in Information Communication

Sanjeeb Kumar Behera[1], Minati Mishra[2]
[1,2]Department of Information and Communication Technology
Fakir Mohan University, Balasore, Odisha
Email: sanjeeb4ever@gmail.com[1], minatiminu@yahoo.com[2]

*Abstract:* **With the growth of communication over computer networks, how to maintain the confidentiality and security of transmitted information have become some of the important issues. In order to transfer data securely to the destination without unwanted disclosure or damage, nature inspired hide and seek tricks such as, cryptography and Steganography are heavily in use. Just like the Chameleon and many other bio-species those change their body color and hide themselves in the background in order to protect them from external attacks, Cryptography and Steganography are techniques those are used to encrypt and hide the secret data inside other media to ensure data security. This paper discusses the concept of a simple spatial domain LSB Steganography that encrypts the secrets using Fibonacci- Lucas transformation, before hiding, for better security.**

## I. Introduction

Data is the backbone of any communication. With the drastic decline in the cost of the computing infrastructures required to copy, print, process and transmit information, modern communication systems make the distribution of documents instant and economic. This has given rise to issues such as illegal copying and distribution of copyrighted materials. As a result, people have started thinking of how to prevent such illicit activities and protect their works. The answer to this search has given birth to the magic concept of Steganography- the technique of message hiding. Nowadays, these techniques have become popular in a number of application areas including few negative areas [1].

The concept of data hiding has come up firstly to protect a piece of information so that data is secured and does not go to unintended destination, secondly, to protect the confidentiality and integrity of data against unauthorized access [2]. Data hiding techniques are broadly classified into four categories such as, Covert channels, Steganography, Anonymity and Copyright marking [3]. Cryptography is another information securing technique where the message is transformed so as to make its meaning obscure to malicious people who try to intercept it [4].

### A. Steganography

Steganography is one of the important and widely used data hiding technique that is used for invisible communication of messages. This is done by hiding information in other information. The term Steganography has been derived from the Greek words "stegein" means "to cover" and "grafein" means "to write". Hence, Steganography can be considered to be a procedure of covered writing [5]. Since last few decades, this approach of information hiding has become popular in a number of application areas.

### B. Types of Steganography

There exist both spatial as well as transform domain image Steganography methods. The transform domain methods are generally used for authentication purposes and the spatial domain methods for Steganography. Depending upon the media used for cover, Steganography can further be classified into text, audio, image or video Steganography. Steganographic methods generally use the redundant bits of the cover medium for information hiding hence text is not very often used as cover since text files provide a small amount of redundancy. Audio/video though provide a lot of redundancy but are very complex to use therefore, image Steganography is widely used. Today, images the most popular cover media used for Steganography where an altered image with some amount of intensity variation remains indistinguishable from the original image to human vision. In this work images are used as covers to hide the secret information. LSB substitution is a popular spatial domain method that replaces the lower order image bits, those do not carry much useful image information, by the secret message bits [6]. Two other technologies closely related to Steganography are watermarking and fingerprinting. These technologies are mainly concerned with authentication and protection of intellectual property [7].

### C. Steganography: A game of hide and seek

If we compare Steganography to the camouflaging behaviors of living species, then we can see that there are some species those naturally protect their lives from external attacks by using several deception tricks. Chameleons, some butterflies and few other bio-species change their body colors depending upon the surrounding and blend them into the surrounding so that the external attackers will not be able know about their presence. Steganographic techniques adopt these tricks from the nature and demands that the cover and Stego media should visually look same so that the external attackers will not know the presence of the secret message.

## II. LSB Based Image Steganography

Images are the most common types of digital media used for Steganography. Digital image have large amount of redundant data and for this reason it is possible to hide message inside image files [1-7]. From mathematical and computer point of view, an image is a collection of numbers that represent the intensities of image points. This numeric representation forms a grid and the individual points are referred to as pixels. Image Steganography is about exploiting the human visual system (HVS) [7][9]. If any specific color is viewed closely it can be observed that single or two three digits modification to the intensity levels are imperceptible to the human eyes i.e pixel with values of (255, 255, 255) to (248, 248, 248) in RGB representation all will be perceived as white by HVS. For example figure.1, 4 different shades of each red and green are shown with values ranging between (127, 0, 0) to (121,0, 0) and (0, 127, 0) to (0, 121,0) but hardly our human visual system can find out any difference between those color sub bands. Spatial domain Steganography exploits this feature of digital images to hide a large amount of data modifying the lower order bit planes of the cover image. The advantage of LSB based data hiding method is, it is simple to embed the bits of the message directly into the LSB plane of image. The LSB modification does not result in image distortion and thus the resulted Stego-image looks identical to the cover image [3].

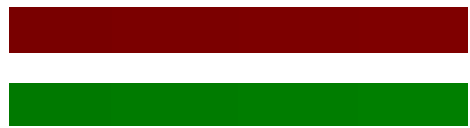

**Figure 1.** Four different shades of Red and Green with intensities ranging between 127 to 121

## III. Proposed Method

To enhance the embedding capacity of image Steganography and to provide better security to the embedded secret, here we purpose a framework that hides the encrypted secret into multiple lower order bit planes. Hiding data using LSB modification alone is not going to be secure hence, to enhance the security level of the embedded secret, Fibonacci-Lucas transformation has been used to encrypt the message before embedding. The combination of both these methodologies gives security, capacity and robustness for secure data transmission in an open channel. In figure.2, is given the block diagram of the overall procedure.

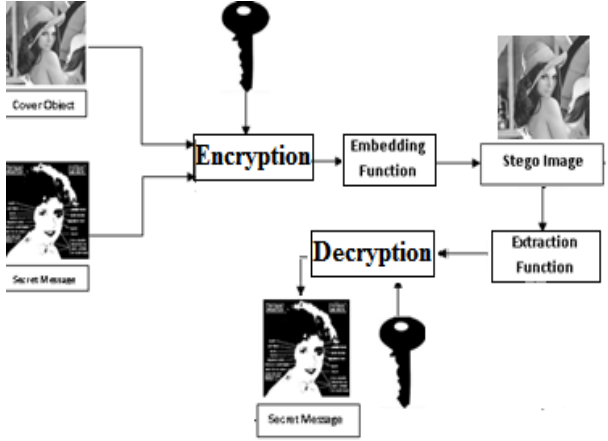

**Figure 2**. Block diagram of Steganography System

A. *Fibonacci-Lucas transformation[8]*

The Fibonacci-Lucas Transform can be defined as the mapping FL: $T^2 \rightarrow T^2$ such that:

$$\begin{pmatrix} x' \\ y' \end{pmatrix} = \begin{pmatrix} F_i & F_{i+1} \\ L_i & L_{i+1} \end{pmatrix} \begin{pmatrix} x \\ y \end{pmatrix} \pmod{N} \quad (1)$$

Where, x, y $\in \{0,1,2,\ldots,N-1\}$, $F_i$ is the $i^{th}$ term of the Fibo 11 series and $L_i$ is $i^{th}$ the term of Lucas series. In table.1 are given some of the terms of *F11* and Lucas series. Denoting $\begin{pmatrix} F_i & F_{i+1} \\ L_i & I_{i+1} \end{pmatrix}$ as $FLT_i$, the first matrix of this series will be given by:

$$F(11)\,LT_1 = \begin{pmatrix} F_1 & F_2 \\ L_1 & L_2 \end{pmatrix} = \begin{pmatrix} 1 & 1 \\ 2 & 1 \end{pmatrix} \quad (2)$$

**Table 1.** Terms of Fibonacci and Lucas Series

| N | 1 | 2 | 3 | 4 | 5 | 6 | 7 | 8 | 9 | 10 | 11 |
|---|---|---|---|---|---|---|---|---|---|----|----|
| $F11_n$ | 1 | 1 | 2 | 3 | 5 | 8 | 13 | 21 | 34 | 55 | 89 |
| $L_n$ | 2 | 1 | 3 | 4 | 7 | 11 | 18 | 29 | 47 | 76 | 123 |

Continuing in this way we can form an infinitely many transforms, all of which are periodic in nature with a maximum possible periodicity of $N^2-1$ and those which produce scrambling patterns different from each other [7].

B. *Proposed Algorithms*

i) Encryption Algo:

INPUT: Cover image *C*, secret messages $S_i$, transformation map $F(11)L_6$, key: $K_{ri}$

*Step1:* Find the period *P* of the map $F(11)L_6$

For all $S_i$ do

*Step2:* Iterate $S_i$ for $P - K_{ri}$ times with the map to get the scrambled secrets $S_i`$

ii) Decryption Algo:

INPUT: Scrambled secrets $S_i`$, transformation map $F(11)L_6$, key: $K_{ri}$

For all $S_i`$

*Step:* Iterate $S_i`$ for $K_{ri}$ times with the map to get back the secrets $S_i$

iii) Embedding Algo:

INPUT: Cover image *C*, scrambled secrets $S_i`$

*Step1:* Bit slice *C*

*Step2:* Replace the lower order bit planes of *C* with $S_i`$

*Step3:* Reconstruct the image with the replaced bit planes to obtain the Stego image *C`*.

iv) Extraction Algo:

INPUT: Stego image *C`*

S*tep1:* Bit slice *C`*

*Step2:* Retrieve the encrypted secrets $S_i`$ from the lower order bit planes

*Step3:* for each $S_i`$, run the decryption algo to get the secrets $S_i$.

## IV. RESULT & ANALYSIS

In this paper we have taken five cover images such as trees.tif, lena.png, cameraman.tif, kids.tif and rice.png and three secret messages S1, S2, S3. First the secret messages are encrypted for 125, 93 and 222 times respectively using the $F(11)L_6$ map i. e. $\begin{pmatrix} 8 & 13 \\ 11 & 18 \end{pmatrix}$ and then the encrypted secrets are embedded into the $1^{st}$, $2^{nd}$ and $3^{rd}$ bit planes from the LSB of the cover images to obtain the Stego images. The secret messages and the encrypted secrets are given in figure 3 whereas the cover and the corresponding Stego images with three bit plane embedding are given in figure 4. It can be seen from figure 4 that the Stego images give no visual clues of hidden secret and these look visually indifferent from the original images. The peak signal to noise ratio (PSNR) and the MSE (mean squared errors) values (given in table 2) also lie within the acceptable range proving no degradation in visual quality of the Stego images. The receivers, to retrieve the secret first, need to bit-slice the Stego images and retrieve the encrypted secrets. It can be seen from figure 3 that the encrypted secrets resemble the noise planes and give no clue of the secret messages.

**Table 2.** PSNR and MSE of various Stego images

| Images | PSNR | | | MSE | | |
|---|---|---|---|---|---|---|
| | I bit Plane | II bit plane | III bit plane | I bit Plane | II bit plane | III bit plane |
| Cameraman | 52.9 | 51.3 | 46.7 | 0.33 | 0.48 | 1.37 |
| Lena.png | 52.9 | 51.2 | 46.7 | 0.33 | 0.49 | 1.40 |
| Trees.tif | 52.5 | 52.3 | 47.2 | 0.36 | 0.39 | 1.23 |
| Kids.tif | 52.9 | 52.2 | 47.3 | 0.33 | 0.39 | 1.20 |
| Rice.png | 52.9 | 51.2 | 46.7 | 0.33 | 0.49 | 1.38 |

At the receiving end, the receiver needs to iterate the encrypted secrets for $K_r$ number of times where, $K_r$ : the receiver's decryption key is equal to $P-K_s$ i.e. the transform period of the map (256 in this case) minus the sender's encryption key. An unauthorized receiver without knowing the key $K_r$ or the map $F(11)L_6$ will not be able the retrieve the secret messages from the scrambled planes hence, this method provides double security against unauthorized access.

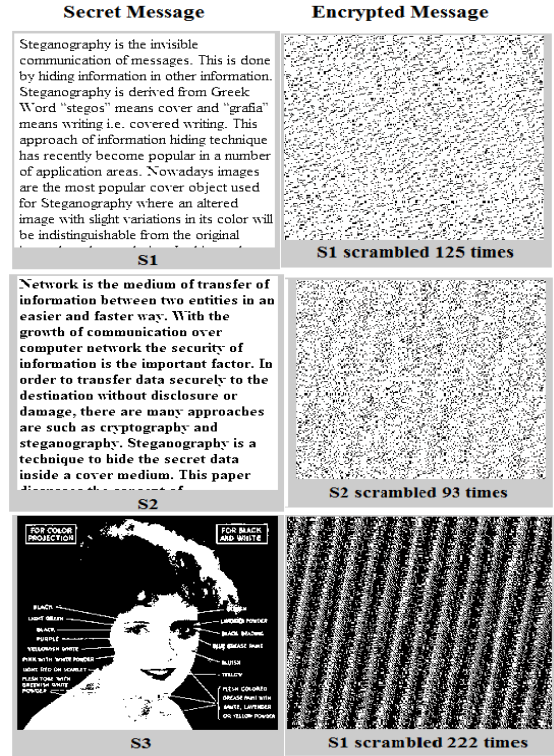

**Figure 3**. Secret messages and their corresponding encrypted messages by F(11)L$_6$ map

## V. CONCLUSION

In spatial domain Steganography, researchers generally use the Arnold Cat map for encryption purpose but that map being a fixed map and being periodic nature so, anyone can get back the secret by iterating the scrambled message systematically for a certain number of times. Hence, Arnold is not considered to be a stronger map [8]. The method discussed in this paper is very easy to implement and at the same time provides higher security to the embedded secret because in this case the map is not a single map. The transmitter and receiver, a priory, can agree to use a certain map out of many possible maps and an adversary, who is not aware of the map that has been for encryption, cannot retrieve the secret. In [10] are suggested still another set of encryption maps and a solution to the key and map transfer problem. In that paper, the authors have also ensured robustness by compromising with the capacity to some extent and by selective embedding of the secret.

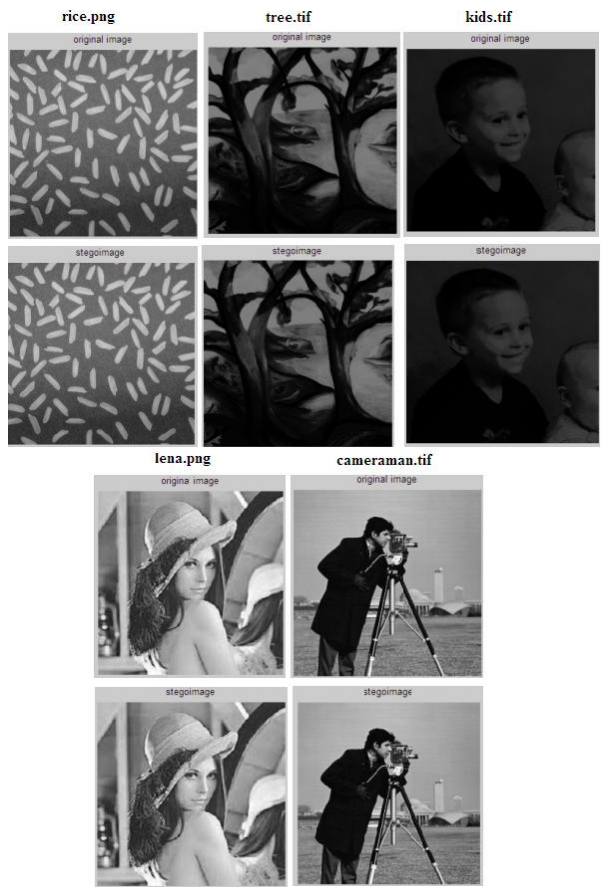

**Figure 4**. Various Cover and Stego images